\documentclass[%
 reprint,
superscriptaddress,
 amsmath,amssymb,
aps,
longbibliography,
nobalancelastpage
]{revtex4-1}

\usepackage{graphicx}
\usepackage{dcolumn}
\usepackage{bm}
\usepackage{hyperref}
\usepackage{siunitx}
\usepackage{todonotes}
\usepackage{txfonts}
\usepackage{xcolor}

\AtBeginDocument{%
    \newwrite\bibnotes
    \def\bibnotesext{Notes.bib}
    \immediate\openout\bibnotes=\jobname\bibnotesext
    \immediate\write\bibnotes{@CONTROL{REVTEX41Control}}
    \immediate\write\bibnotes{@CONTROL{%
    apsrev41Control,author="08",editor="1",pages="1",title="0",year="1"}}
     \if@filesw
     \immediate\write\@auxout{\string\citation{apsrev41Control}}%
    \fi
}%

\begin{document}


\title{Frequency fluctuations in nanomechanical silicon nitride string resonators}
\author{Pedram Sadeghi}
\affiliation{Institute of Sensor and Actuator Systems, TU Wien, 1040 Vienna, Austria}
\author{Alper Demir}
\affiliation{Department of Electrical Engineering, Ko\c{c} University, Istanbul 34450, Turkey}
\author{Luis Guillermo Villanueva}
\affiliation{EPFL-STI-IGM-NEMS, Lausanne, Switzerland}
\author{Hendrik K{\"a}hler}
\affiliation{Institute of Sensor and Actuator Systems, TU Wien, 1040 Vienna, Austria}
\author{Silvan Schmid}
 \email{silvan.schmid@tuwien.ac.at}
\affiliation{Institute of Sensor and Actuator Systems, TU Wien, 1040 Vienna, Austria}
\date{\today}

\date{\today}

\begin{abstract}

High quality factor ($Q$) nanomechanical resonators have received a lot of attention for sensor applications with unprecedented sensitivity. Despite the large interest, few investigations into the frequency stability of high-$Q$ resonators have been reported. Such resonators are characterized by a linewidth significantly smaller than typically employed measurement bandwidths, which is the opposite regime to what is normally considered for sensors. Here, the frequency stability of high-$Q$ silicon nitride string resonators is investigated both in open-loop and closed-loop configurations. The stability is here characterized using the Allan deviation. For open-loop tracking, it is found that the Allan deviation gets separated into two regimes, one limited by the thermomechanical noise of the resonator and the other by the detection noise of the optical transduction system. The point of transition between the two regimes is the resonator response time, which can be shown to have a linear dependence on $Q$. Laser power fluctuations from the optical readout is found to present a fundamental limit to the frequency stability. Finally, for closed-loop measurements, the response time is shown to no longer be intrinsically limited but instead given by the bandwidth of the closed-loop tracking system. Computed Allan deviations based on theory are given as well and found to agree well with the measurements. These results are of importance for the understanding of fundamental limitations of high-$Q$ resonators and their application as high performance sensors.

\end{abstract}

\keywords{}

\maketitle

\section{Introduction}

As the size of a mechanical resonator is minimized, the responsivity towards changes in the resonator environment is increased \cite{Schmid2016,Ekinci2005}. This has led to a tremendous amount of research in the field of nanomechanical resonators, with devices being employed to detect mass \cite{Ekinci2004_1,Naik2009,Chaste2012}, force \cite{Kozinsky2006,Moser2013}, and temperature \cite{Pandey2010,Larsen2011,Zhang2013,Piller2019}. Most studies have employed changes in the resonance frequency $f_\text{0}$ as the detection principle.

For sensing applications, the minimum detectable frequency shift $\delta f$ is determined by the precision of the resonance frequency measurement scheme. Quantifying any noise that presents a limit to the frequency stability of a resonator is thus crucial when designing sensors. Various intrinsic noise sources influence the stability of a device, including thermomechanical noise \cite{Ekinci2004_2}, adsorption-desorption noise \cite{Vig1999}, defect motion \cite{Cleland2002}, surface diffusion \cite{Atalaya2011}, and damping fluctuations \cite{Maillet2018}. If the underlying limiting noise source is white, the average relative frequency noise can be quantified by \cite{Ekinci2004_2,Robins1984}
\begin{equation}\label{eq:robin}
    \left\langle\frac{\delta f}{f_\text{0}}\right\rangle \approx \frac{1}{2 Q} \frac{1}{\text{SNR}},
\end{equation}
where $Q$ is the quality factor and SNR is the signal-to-noise ratio (ratio of the maximum driven signal to the noise floor). From equation~\ref{eq:robin}, it is apparent that minimizing the system noise results in optimal frequency stability. This would imply that thermomechanical noise, arising from a coupling between the resonator and a thermal bath of randomly distributed phonons, presents an intrinsic limit to the frequency stability in the regime where it can be resolved \cite{Albrecht1991,Cleland2005,Gavartin2013}. 

An important factor when measuring the displacement of a nanomechanical resonator is the relation between the resolution bandwidth (BW) of the readout electronics, typically on the order of kHz, and the resonance linewidth ($\Gamma$) of the resonator. Most fundamental studies thus far have operated in the regime where $\text{BW} \ll \Gamma$ \cite{Sansa2016}. For high-$Q$ resonators, $\Gamma$s on the order of 1 Hz are routinely achieved and can even be on the order of a few mHz \cite{Tsaturyan2017,Ghadimi2018}. Various studies have been published investigating the frequency stability of high-$Q$ resonators with different conclusions. Fong et al. investigated the frequency and phase noise of high-$Q$ silicon nitride resonators, and found that frequency fluctuations were a result of defect motion with a broad distribution of relaxation times \cite{Fong2012}. Furthermore, it was concluded that increasing $Q$ results in an increased susceptibility to the intrinsic frequency fluctuations in the sample. Roy et al. reported enhanced frequency stability in resonators made from silicon, when $Q$ is lowered as the $\text{SNR}$ is increased, operating at the onset of Duffing nonlinearity in a thermomechanically limited regime \cite{Roy2018}. This was attributed to a flattening of the phase noise spectrum at low frequencies. A thorough theoretical investigation of the same scenario by Demir et al. however found no $Q$-dependence of the frequency stability \cite{Demir2020}.

In this report, we investigate the frequency stability of high-$Q$ nanomechanical string resonators made from silicon nitride (SiN). Such resonators routinely achieve large $Q$s as a result of the dissipation dilution effect \cite{Verbridge2006,Schmid2011,Unterreithmeier2010}. All measurements are performed in the thermomechanically resolved regime and BWs on the order of a kHz are chosen similar to previous investigations. In the first part, open-loop tracking of the frequency is performed using the phase of the resonator and the Allan deviation is employed to quantify the frequency noise. As a result of the setup with $\text{BW}\gg  \Gamma$, both thermomechanical noise and detection noise of the measurement system are captured in the Allan deviation. The two regimes are separated at the integration time $\tau = \tau_\text{r} = 1/\left(\pi \Gamma\right)$, which is the response time of the resonator. 

We find that for lower vibrational amplitudes, the resonator is always thermomechanically limited. At larger amplitudes, however, the laser power fluctuations of the optical readout presents a lower limit to the frequency stability. A systematic reduction of $Q$ results in faster response times, but the lower limit from the laser remains regardless of the $Q$. Finally, closed-loop measurements are performed, where it can be shown that $\tau_{\text{r}}$ can be increased through variation of the loop bandwidth. All Allan deviation measurements are corroborated by computations based on a theoretical model \cite{Demir2020} and good agreement is observed both for open-loop and closed-loop.

\section{Methods}

\subsection{Measurement setup}

\begin{figure}
  \centering
  \includegraphics[width=0.45\textwidth]{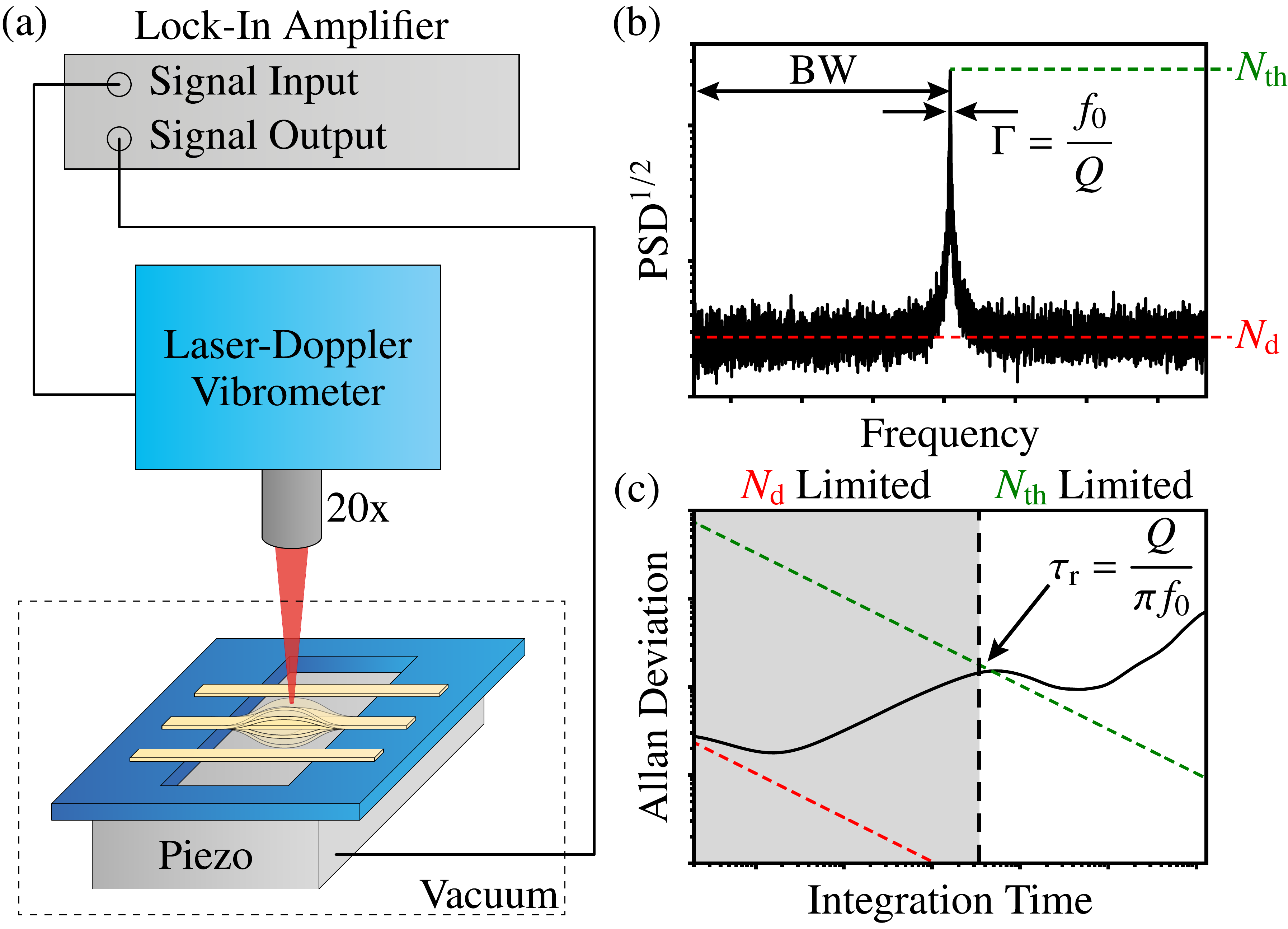}
  \caption[Figure Setup]{(a) Schematic of the measurement setup. Mechanical vibrations of the string resonator are measured using a commercial laser-Doppler vibrometer. The sample is placed inside a vacuum chamber equipped with a needle valve for pressure control and resonances are actuated using a piezo. (b) Exemplary plot of the square root of the power spectral density $\text{PSD}^{1/2}$ versus frequency around the fundamental resonance, highlighting the two sources of noise: Thermomechanical noise $N_{\text{th}}$ and detection noise $N_{\text{d}}$. The resonance linewidth $\Gamma$ and the resolution bandwidth BW are shown as well. (c) Allan deviation curve showing the limits presented by both noise sources along with the resonator response time $\tau_\text{r} = 1/\left(\pi \Gamma\right) = Q/\left(\pi f_\text{0}\right)$. For longer integration times ($\tau > \tau_{\text{r}}$), the Allan deviation is limited by thermomechanical noise (green dashed line) and thermal drift, while detection noise (red dashed line) limits the stability at shorter integration times ($\tau < \tau_{\text{r}}$).}\label{fig:setup}
\end{figure}

The fabrication process of the nanomechanical string resonators is provided elsewhere \cite{Schmid2010}. A schematic of the measurement setup is given in Fig.~\ref{fig:setup}(a). Resonators are actuated using a piezo (NAC2003 from CTS Corporation or a custom-sized plate of piezo ceramic type PIC252 from PI Ceramic GmbH) and the out-of-plane motion detected using a laser-Doppler vibrometer (MSA-500 from Polytec GmbH) equipped with a helium-neon laser ($\lambda = 633$~nm). All measurements are performed in a vacuum chamber at various pressures, controlled using a needle valve (8LVM-10KF-VV-A from VACOM). The analog signal of the vibrometer is sent to a lock-in amplifier (HF2LI from Zurich Instruments) equipped with a phase-locked loop (PLL), where the frequency stability is investigated.

\subsection{Allan deviation}

In this report, the frequency stability is quantified through the Allan deviation, defined as \cite{Allan1966}:
\begin{equation}\label{eq:allan_dev}
    \sigma_{\text{A}}\left(\tau\right) = \sqrt{\frac{1}{2\left(N-1\right)}\sum_{i=1}^{N-1}\left(\frac{\overline{f}_{i+1}-\overline{f}_{i}}{f_0}\right)^2},
\end{equation}
where $N$ is the number of samples of the resonance frequency $\overline{f}_{1}...\overline{f}_{N}$, each averaged over an integration time $\tau$. Measurements of the frequency fluctuations are made either in an open-loop or closed-loop configuration. For open-loop measurements, the sample is actuated at the resonance frequency and the phase $\phi\left(t\right)$ is recorded for a certain amount of time. Conversion from phase to frequency is then made using the phase response of the resonator. In the linear driving regime, the phase shift $\Delta \phi$ caused by a frequency shift $\Delta f$ close to resonance is given as $\Delta\phi  \approx \left(2 Q/f_\text{0}\right) \Delta f$ \cite{Schmid2016}. $Q$s of the resonators are measured using the ring-down method. In the closed-loop configuration, the PLL of the lock-in amplifier is employed to track the frequency $f\left(t\right)$ directly. Various harmonics appearing in the recorded signals, either from the mains supply ($<$100~Hz), turbomolecular pump ($\sim$1500~Hz), and aliasing ($>$2000~Hz), are filtered out during post-processing.

\begin{figure*}
  \centering
  \includegraphics[width=\textwidth]{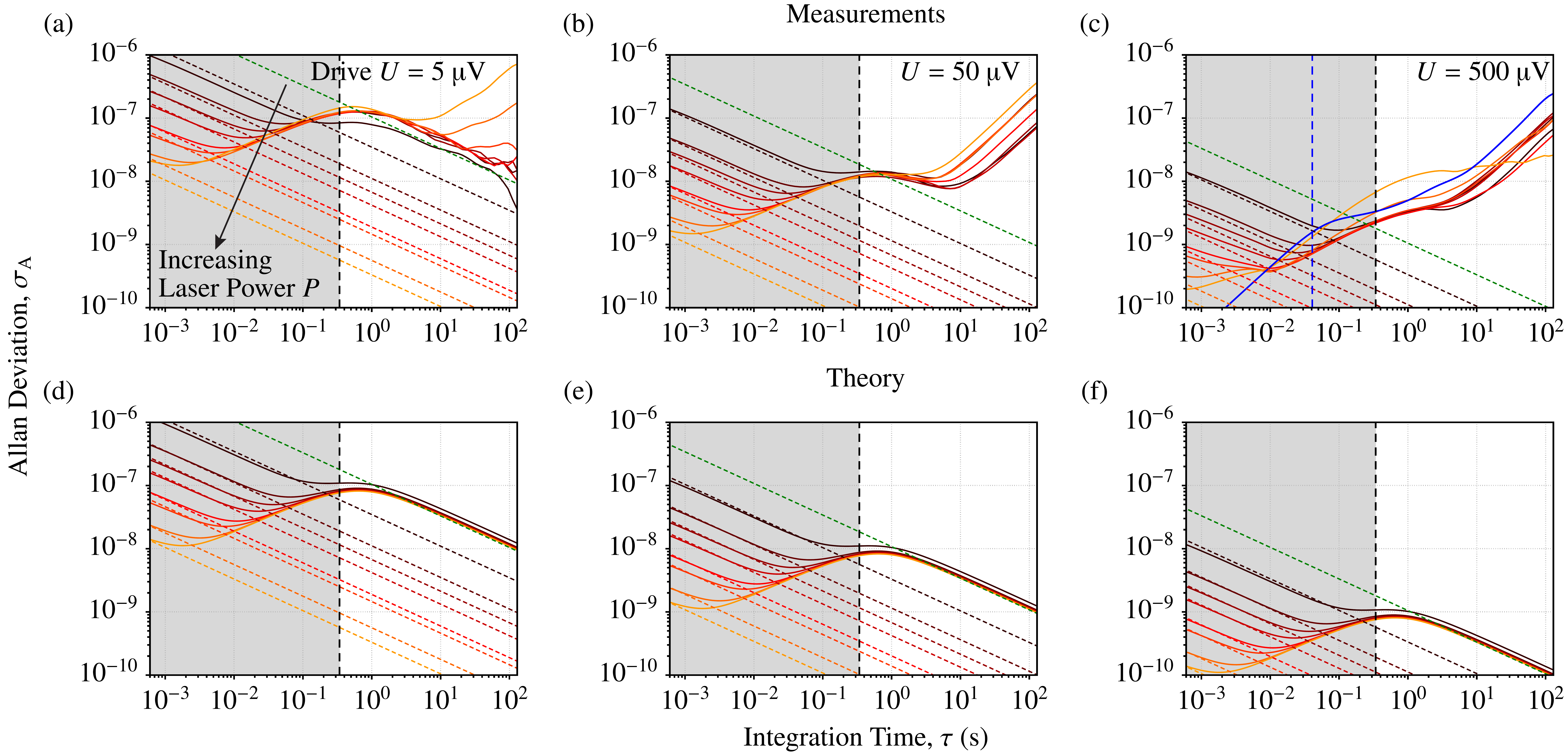}
  \caption[Figure Laser]{Effect of varying detection noise on the Allan deviation for open-loop tracking. Experimental Allan deviations were recorded by varying the vibrometer laser power $P$ between 1.5 \si{\micro\watt} to 270 \si{\micro\watt}, while setting the piezo driving voltage $U = 5$~\si{\micro\volt} (a), $U = 50$~\si{\micro\volt} (b), and $U = 500$~\si{\micro\volt} (c). Corresponding simulated Allan deviations are shown in (d)-(f). Calculations using equation \ref{eq:allan_white} are shown as well with $N_{\text{l}} = N_{\text{th}}$ (green dashed lines) and $N_{\text{l}} = N_{\text{d}}$ (red-colored dashed lines). Black vertical dashed lines highlight $\tau = \tau_\text{r}$. In (c), an Allan deviation curve corresponding to laser power fluctuations for $P = 270$~\si{\micro\watt} is given as well (blue line), which is low-pass filtered during post-processing at a frequency corresponding to the resonator thermal time constant $\tau_\text{th}$ (blue vertical dashed line).}\label{fig:laser}
\end{figure*}

Theory based Allan deviation curves are generated by first computing the spectrum of fractional frequency fluctuations, $S_\text{y}\left(f\right)$, using experimental data for $f_\text{0}$, $Q$, thermomechanical and detection noise levels, driven vibrational amplitude of the resonator and the demodulation filter characteristics of the lock-in amplifier, in conjunction with a theoretical model~\cite{Demir2020}. Then, the Allan variance (the Allan deviation $\sigma_{\text{A}}$ squared) can be estimated through numerical evaluation of the following integral:
\begin{equation}\label{eq:allan_integral}
    \sigma^2_{\text{A}}\left(\tau\right) = \frac{2}{\pi^2 \tau^2}\int_{-\infty}^{+\infty} \frac{\left[\text{sin}\left(\pi \tau f\right)\right]^4}{f^2} S_\text{y}\left(f\right) \text{d}f.
\end{equation}

\subsection{Sources of noise}

An exemplary scan of the noise of a resonator over a region of size twice the BW is shown in Fig.~\ref{fig:setup}(b). From this plot, the two noise sources can be defined: (1) Thermomechanical noise from the resonator, $N_{\text{th}}$, and (2) background, or detection, noise, $N_{\text{d}}$. How these noise sources manifest themselves in the Allan deviation can be seen in Fig.~\ref{fig:setup}(c). For this purpose, we define the resonator response time, or time constant, $\tau_{\text{r}}$. For integration times $\tau > \tau_{\text{r}}$, $\sigma_{\text{A}}$ is mainly limited by thermomechanical noise (white-shaded region in Fig.~\ref{fig:setup}(c)). At larger $\tau$, however, thermal drift typically increases $\sigma_{\text{A}}$. When $\tau < \tau_{\text{r}}$, $\sigma_{\text{A}}$ falls continuously until limited by background noise of the system (grey-shaded region in Fig.~\ref{fig:setup}(c)).

The investigated string sample is of length $L = 1000$~\si{\micro\meter}, width $w = 6$~\si{\micro\meter}, thickness $h = 312$~nm, tensile stress $\sigma = 0.85$~GPa, and $Q$ on the order of $10^5$ at high vacuum. Only the fundamental out-of-plane mode is considered, which has a resonance frequency $f_0 = 264$~kHz. The BW was set to 3598~Hz and the sampling rate to 28784~Sa/s. The SNR is controlled through variation of the vibrometer laser power $P$, which decreases detection noise for increasing $P$, and the piezo driving voltage $U$, which increases the vibrational amplitude, or signal, for larger $U$. For the integral given in equation~\ref{eq:allan_integral}, the white noise asymptote for the Allan deviation can be shown to be \cite{Demir2020}: \begin{equation}\label{eq:allan_white}
    \sigma_{\text{A}}\left(\tau\right) = \frac{1}{2 Q}\frac{N_{\text{l}}}{A}\sqrt{\frac{1}{\tau}}.
\end{equation}
Here, $N_{\text{l}}$ is the noise level, defined as the square root of the one-sided noise spectral density (in units of $\text{X}/\sqrt{\text{Hz}}$) resulting from thermomechanical ($N_{\text{th}}$) or detection ($N_{\text{d}}$) noise and $A$ is the peak driven amplitude (in $\text{X}$). $\text{X}$ represents units of voltage, displacement, etc. 
In the case of thermomechanically limited noise, the noise level (in $\text{m}/\sqrt{\text{Hz}}$) is analytically defined as \cite{Schmid2016}:
\begin{equation}\label{eq:thermo}
    N_{\text{th}} = \sqrt{\frac{k_{\text{B}} T Q}{2 \pi^3 m_{\text{eff}} f_{\text{0}}^3}},
\end{equation}
where $k_{\text{B}}$ is the Boltzmann constant, $T$ is the temperature, and $m_{\text{eff}}$ is the effective mass of the resonator. Calculations of the Allan deviation based on equation~\ref{eq:allan_white} are shown in Fig.~\ref{fig:setup}(c) as well.

\section{Results and Discussion}

\subsection{Influence of signal-to-noise ratio in open-loop configuration}

\begin{figure}
  \centering
  \includegraphics[width=0.45\textwidth]{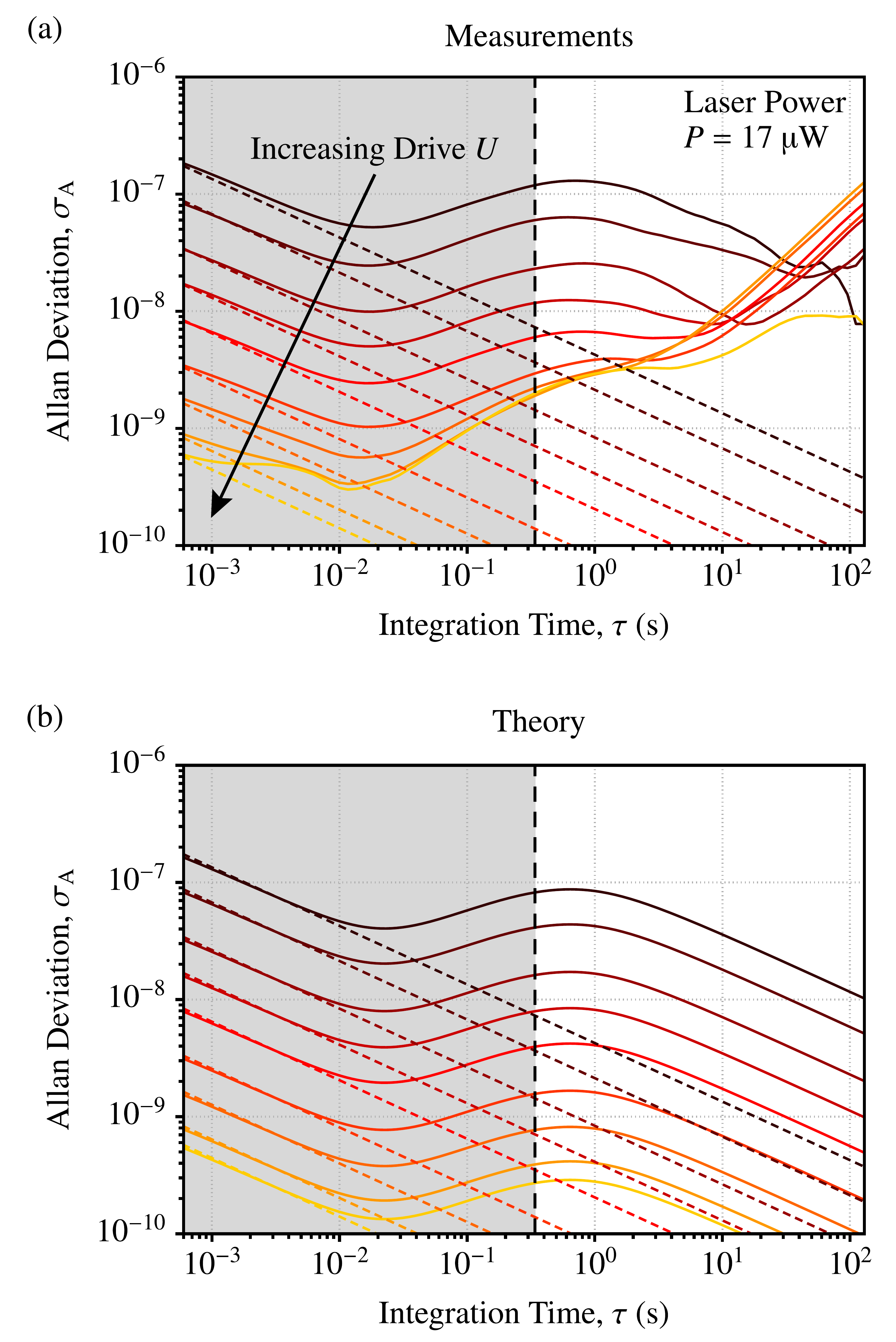}
  \caption[Figure Drive]{Effect of varying vibrational amplitude on the Allan deviation for open-loop tracking. (a) Measured Allan deviations for varying piezo driving voltages from $U = 5$~\si{\micro\volt} to $U = 1.5$~\si{\milli\volt}. (b) Simulated Allan deviations for the data presented in (a). Red-colored dashed lines are detection noise limits calculated using equation \ref{eq:allan_white}. Black vertical dashed lines highlight $\tau = \tau_{\text{r}}$.}\label{fig:drive}
\end{figure}

First, the influence of SNR on the Allan deviation is studied. On the one hand, the detection noise level is controlled by tuning the laser power of the optical readout. On the other hand, the vibrational amplitude, the signal, of the nanomechanical string is controlled via the piezo driving voltage. Fig.~\ref{fig:laser} presents measured and theory-based $\sigma_{\text{A}}$'s for various SNRs. The vibrational amplitudes measured in the electrical domain are converted from volts to metres using the decoders of the vibrometer, and used in theory-based computations of the Allan deviation. In addition, $\sigma_{\text{A}}$ values calculated using equation~\ref{eq:allan_white} are presented as well both for detection noise limits (red-colored dashed lines) and the thermomechanical limit (green dashed line). The vertical dashed lines highlight $\tau = \tau_{\text{r}} = Q/\left(\pi f_\text{0}\right) = 0.34$~s. These measurements are performed under a high vacuum ($p \leq 1 \cdot 10^{-5}$~mbar) in order to keep the $Q$ maximized. In each individual plot, $U$ is kept fixed, while $P$ is varied from 1.5~\si{\micro\watt} to 270~\si{\micro\watt}.

Fig.~\ref{fig:laser}(a)-(c) shows measured $\sigma_{\text{A}}$ curves for the cases of $U = 5$ \si{\micro}V, $U = 50$~\si{\micro\volt}, and $U = 500$~\si{\micro\volt}. For the first case, $U = 5$~\si{\micro\volt}, the resonator is well in the linear regime. The laser power can be observed to decrease $\sigma_{\text{A}}$ in the case of $\tau < \tau_{\text{r}}$, which is the regime limited by detection noise, while the thermomechanically limited regime ($\tau > \tau_{\text{r}}$) displays no reduction. This is to be expected, as only the detection noise gets altered by the laser power. Thermal drifts appear to play a larger role for higher laser powers, which will be discussed further below.

Upon increasing the actuation voltage to $U = 50$ \si{\micro}V, an order of magnitude reduction of $\sigma_{\text{A}}$ is observed, indicating that the sample is still in the linear regime. Finally, in the case of $U = 500$~\si{\micro\volt}, the decrease in $\sigma_{\text{A}}$ is no longer linear in the $\tau > \tau_{\text{r}}$ regime. In this case, the Allan deviation is no longer thermomechanically limited and appears to deteriorate with increasing laser power. Since the sample was still driven in the linear regime, Duffing nonlinearity can be ruled out as a limit to frequency stability in our measurements. Instead, the limit is attributed to laser power fluctuations resulting from the optical readout.

An additional $\sigma_{\text{A}}$ measurement (blue curve) representing the frequency noise resulting from laser power fluctuations is plotted in Fig.~\ref{fig:laser}(c). This curve is generated by first measuring the laser power over time with similar demodulation settings as for the other curves. The particular measurement shown in Fig.~\ref{fig:laser}(c) was for a laser power $P = 270$~\si{\micro\watt}. Then, the laser power fluctuations, $\delta P$, can be converted to frequency fluctuations using the relative thermal responsivity, $\delta R$, of the string using the relation: $\langle\delta f/f_0\rangle = \delta R \cdot \delta P$. Based on the shift of the resonance frequency versus laser power, a $\delta R = -0.08$~\si{\watt}$^{-1}$ can be extracted. The frequency response of the resonator to laser power noise is based on the photothermal heating \cite{Larsen2013}, and the string resonator acts as a low-pass filter with a time constant equal to the thermal time constant $\tau_{\text{th}}$. Hence, during post-processing, the raw frequency fluctuations data is low-pass filtered with a pass frequency $f_\text{pass} = 1/(2 \pi \tau_{\text{th}})$. The thermal time constant was measured at high vacuum in a similar way as described by Piller et al. \cite{Piller2020}. A blue vertical dashed line in Fig.~\ref{fig:laser}(c) shows $\tau = \tau_{\text{th}}$ for clarity. The blue Allan deviation curve due to laser noise explains the increasing slopes for longer $\tau$'s well. The slight offset can be attributed to sample or laser aging due to the fact that the laser noise was measured two years after the Allan deviation data was measured. These findings strongly support that laser power fluctuations cause the limit in Allan deviation observed in this study.


Theory based computations corresponding to the measured data presented above can be seen in Figs.~\ref{fig:laser}(d)-(f) \cite{Demir2020}. The effect of reducing detection noise is well reproduced by the theory. Since laser power fluctuations are not captured in the model, thermal drift effects cannot be observed in the computed curves. Hence, in the case of $U = 500$~\si{\micro\volt}, the frequency stability for $\tau > \tau_{\text{r}}$ is still thermomechanically limited for the model. 

The effect of varying piezo drive from $U = 5$~\si{\micro\volt} to $U = 1.5$~\si{\milli\volt} for a fixed laser power $P = 17$~\si{\micro\watt} can be seen in Fig.~\ref{fig:drive}(a). In this case, $\sigma_{\text{A}}$ is improved for all $\tau$-values, since the noise remains unchanged, while the driven signal increases. Initially, a linear dependence on driving amplitude is observed, but for larger values of $U$, $\sigma_{\text{A}}$ reaches a limit in the thermomechanical regime and does not reduce further with increasing $U$. This hard limit again is attributed to the laser power fluctuations as observed in Fig.~\ref{fig:laser}(c). 
Computed Allan deviation curves based on theory are given in Fig.~\ref{fig:drive}(b), displaying good agreement for lower $U$-values, but deviate for larger values due to the fact that the model does not incorporate laser power fluctuations.

From the data presented in Fig.~\ref{fig:laser} and Fig.~\ref{fig:drive}, some comments can be made on the Allan deviation in open-loop configuration. In general, thermomechanical noise presents a fundamental limit to the frequency stability of nanomechanical resonators. Even though the Allan deviation for $\tau < \tau_{\text{r}}$ is limited by detection noise, possibly below the thermomechanical limit, this does not represent the actual performance limit of the resonant sensor. The resonator itself acts as a filter on the phase response and thus $\sigma_\text{A}$ is reduced as a result for $\tau < \tau_{\text{r}}$. However, the resonator phase can not respond to resonance frequency shifts at those short time scales \cite{Demir2020}. Finally, the limit of $\sigma_{\text{A}}$ appearing for larger drive amplitudes is argued to be a result of the optical readout, as discussed in Fig.~\ref{fig:laser}(c).

\subsection{Influence of quality factor in open-loop configuration}

\begin{figure}
  \centering
  \includegraphics[width=0.45\textwidth]{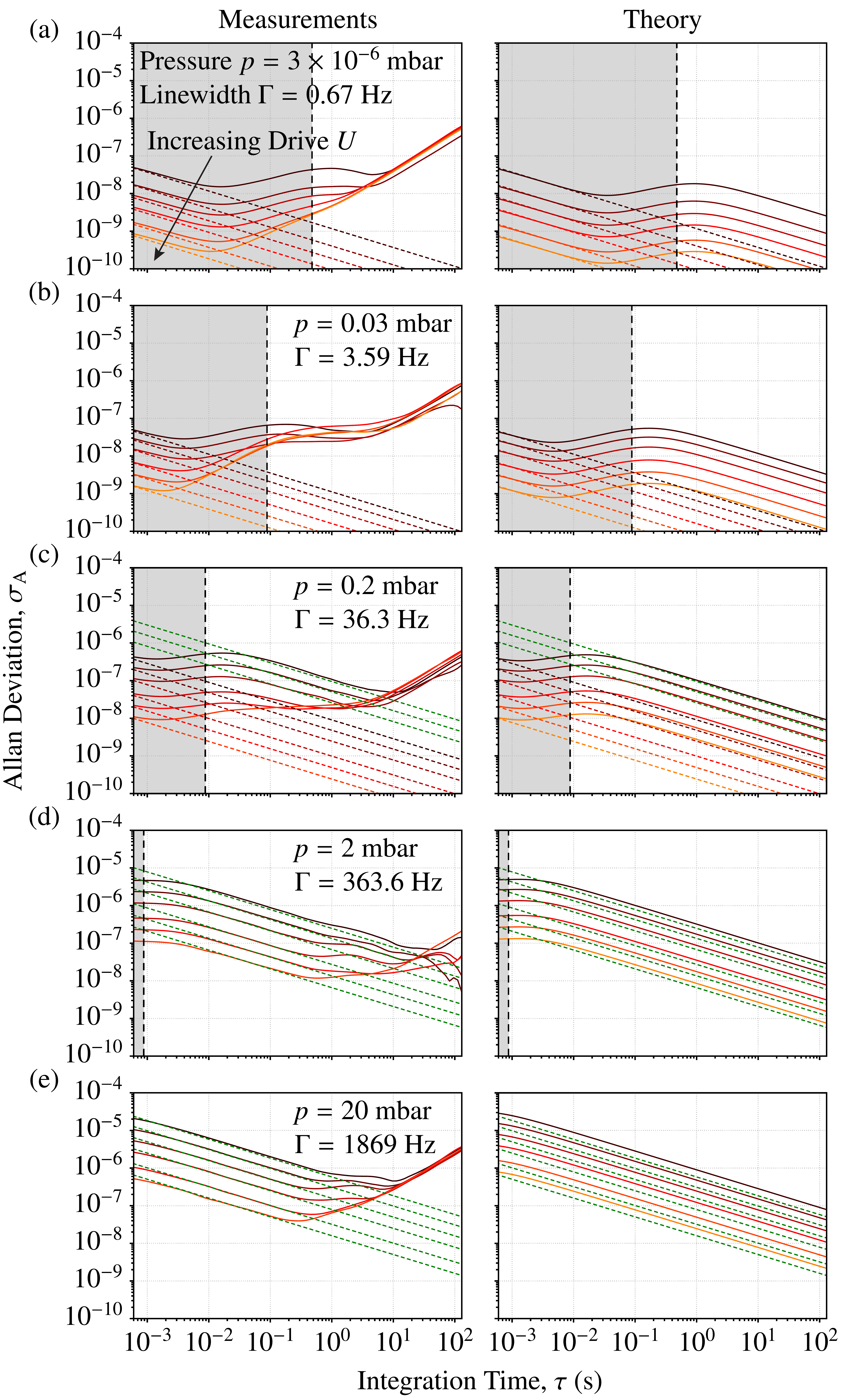}
  \caption[Figure Q]{$Q$-dependence of the frequency stability in open-loop configuration. Measured (left column) and theory based (right column) Allan deviations for resonance linewidths of 0.67~Hz (a), 3.59~Hz (b), 36.3~Hz (c), 363.6~Hz (d), and 1869~Hz (e). Green and red-colored dashed lines are calculations using equation \ref{eq:allan_white} for thermomechanical and detection noise limits, respectively. Black vertical dashed lines highlight $\tau = \tau_{\text{r}}$.}\label{fig:Q}
\end{figure}

The next step is to investigate the influence of the quality factor on the frequency stability of the resonators in open-loop configuration. Using a needle valve, the pressure $p$ in the vacuum chamber is increased from $3 \times 10^{-6}$~mbar up to 20~mbar, which reduces the quality factor due to gas damping \cite{Verbridge2008_2}. Measured and simulated $\sigma_{\text{A}}$ curves for various pressures are shown in Fig.~\ref{fig:Q}. Each curve is recorded for a fixed laser power and varying drive amplitudes. As a result of large drifts in the power of the laser over time, a laser power $P = 17$~\si{\micro}W was used for the data at $p = 3 \times 10^{-6}$~mbar, while the maximum available $P = 270$~\si{\micro\watt} was employed at all other pressures, in order to ensure that the thermomechanical noise could be resolved. On each measured plot, the pressure at which the measurement was recorded is given, along with the linewidth of the resonance.

Experimental $\sigma_\text{A}$'s are shown in the left column of Figs.~\ref{fig:Q}(a)-(e), while theoretical curves are shown in the right column. A clear shift in the transition region between thermomechanical and detection noise limits to lower integration times can be observed. The point of transition consistently lies around $\tau = \tau_{\text{r}}$ and a linear dependence on $Q$ is found for this transition time, as is expected from the definition of $\tau_{\text{r}}$. Furthermore, as $\Gamma$ approaches the BW, the frequency stability is thermomechanically limited for all $\tau$ (neglecting thermal drift). These findings indicate that the response time of a resonator is fundamentally limited by the $Q$ in open-loop configuration. Similar response times can be found by investigating the transient response of the phase in open-loop, as presented in Supplementary Fig.~\ref{fig:time}. For the measured data shown in Figs.~\ref{fig:Q}(a)-(e), $\sigma_\text{A}$ again is limited for long integration times due to laser power fluctuations as discussed above.

\subsection{Closed-loop tracking}

\begin{figure}[ht!]
  \centering
  \includegraphics[width=0.45\textwidth]{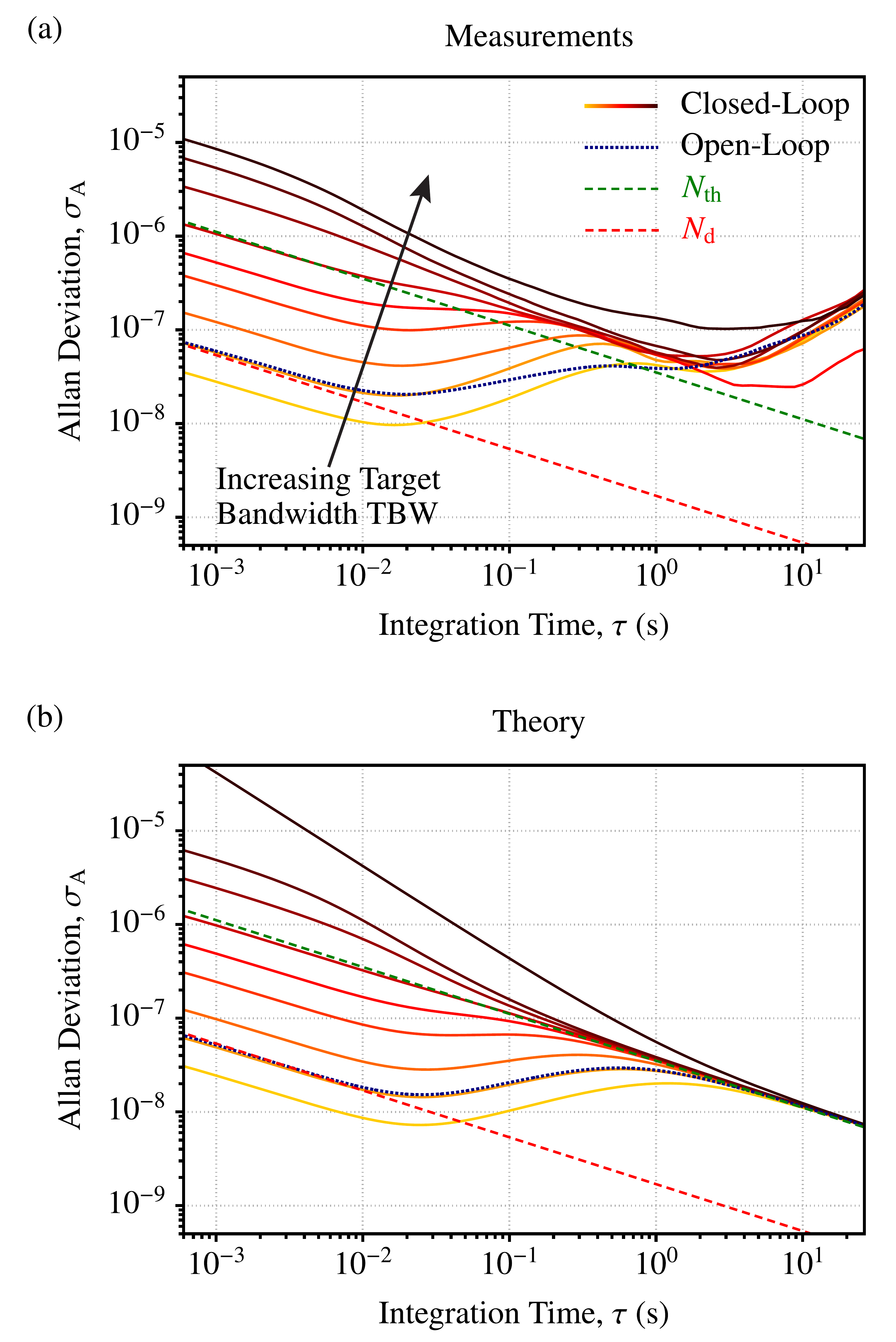}
  \caption[Figure PLL]{Influence of varying phase-locked loop target bandwidth on the frequency stability. (a) Measured and (b) theory based Allan deviations for target bandwidths of (from bottom to top): 0.25~Hz, 0.5~Hz, 1~Hz, 2.5~Hz, 5~Hz, 10~Hz, 25~Hz, 50~Hz, and 719.6~Hz. Blue dotted line is an open-loop measurement. Dashed lines are calculations using equation \ref{eq:allan_white} for thermomechanical (green) and detection (red) noise limits.}\label{fig:PLL}
\end{figure}

Investigation into the limits of frequency stability and response time for closed-loop tracking have also been performed. For these experiments, a PLL is employed to track the frequency. For the lock-in amplifier used here, an additional BW, i.e. the PLL loop bandwidth \cite{Demir2020}, needs to be set, which is called the target BW (from here on called TBW in order to distinguish it from the filter BW). This is the bandwidth for the entire closed-loop system, and typically smaller than the BW of the filter in the demodulator. TBW can be set to a desired value by choosing the controller parameters appropriately \cite{Demir2020}. The effect of varying TBW from 0.25~Hz to 719.6~Hz on $\sigma_\text{A}$ is shown based on measurements and theory in Figs.~\ref{fig:PLL}(a) and (b), respectively. Similar to the open-loop measurements, a demodulator filter BW of 3598~Hz is chosen, while the sampling rate is set to 28784~Sa/s. A $\sigma_\text{A}$ curve recorded in open-loop is shown as well for comparison. In addition, calculations using equation \ref{eq:allan_white} are given both for thermomechanical and detection noise limits.

From the closed-loop data, it can be observed that the closed-loop response time is determined by the TBW of the system instead of the intrinsic response time of the resonator, as opposed to the open-loop measurements. As the TBW is increased, the point of transition between thermomechanical and detection noise shifts to lower $\tau$-values. An additional effect of the PLL is the change in detection noise for different TBWs, e.g. for the smallest TBW satisfying $\text{TBW}<\Gamma$, the detection noise limit is even lower than the one in the open-loop case. With $\text{TBW}<\Gamma$, the filtering provided by the loop, in addition to the filtering by the intrinsic response of the resonator and the demodulator filter, reduces the $\sigma_\text{A}$ of the system. However, in this case, the response time of the system is even slower than the intrinsic $\tau_\text{r}$ of the resonator.

Similar to the open-loop Allan deviations, the smaller $\sigma_\text{A}$ values for $\tau$ less than the system response time are merely resulting from the chosen bandwidths for the system relative to the linewidth of the resonator. As evident from the data in Fig.~\ref{fig:PLL}, $\sigma_\text{A}$ remains thermomechanically limited for larger $\tau$, regardless of the TBW. If frequency shifts are happening at faster time scales in the detection noise limited regime, the loop will not be able to respond as fast. By increasing the TBW beyond $\Gamma$ and much further, the system will be able to follow faster changes, but at the expense of increased noise.

\section{Conclusion}

We have presented experimental and theoretical results for the limitations in frequency stability of high-$Q$ nanomechanical SiN resonators both in open-loop and closed-loop configurations. Through open-loop tracking of the resonator frequency, it was found that the Allan deviation gets separated into two regimes, one limited by thermomechanical noise of the resonator and the other by detection noise of the optical transduction scheme. These regimes could be clearly observed and quantified through variation of both the detection noise floor and the vibrational amplitude. The point of transition between the regimes could be characterized as the response time, $\tau_\text{r}$, of the resonator in the open-loop case. Through variation of $Q$, a linear dependence of $\tau_\text{r}$ on $Q$ could be observed. Furthermore, for the open-loop measurements, laser power fluctuations resulting from the optical readout presented a lower limit to the Allan deviation at higher integration times. Finally, closed-loop tracking was investigated experimentally and theoretically, and it was shown that the closed-loop scheme can be tailored to offer a faster response at the expense of increased noise, when compared to the intrinsic limits of open-loop tracking. Computed Allan deviations based on a theoretical model for both open-loop and closed-loop showed good agreement with measurements. The findings presented here show the limitations of open-loop measurements of the frequency fluctuations and that the method is inadequate for assessing the frequency stability of high-$Q$ resonators. Given that the presented results are not limited to SiN strings but applicable to all types of high-$Q$ resonators, they are of interest for future designs of nanomechanical sensors and fundamental investigations into the ultimate frequency fluctuation limits of such resonators in general.

\begin{acknowledgments}
The authors would like to thank Miao-Hsuan Chien and Markus Piller for many helpful discussions. This work is supported by the European Research Council under the European Unions Horizon 2020 research and innovation program (Grant Agreement-716087-PLASMECS).

\end{acknowledgments}

\appendix
\renewcommand{\thefigure}{S\arabic{figure}}
\setcounter{figure}{0}

\section{Transient response of phase}

\begin{figure}[ht!]
  \centering
  \includegraphics[width=0.45\textwidth]{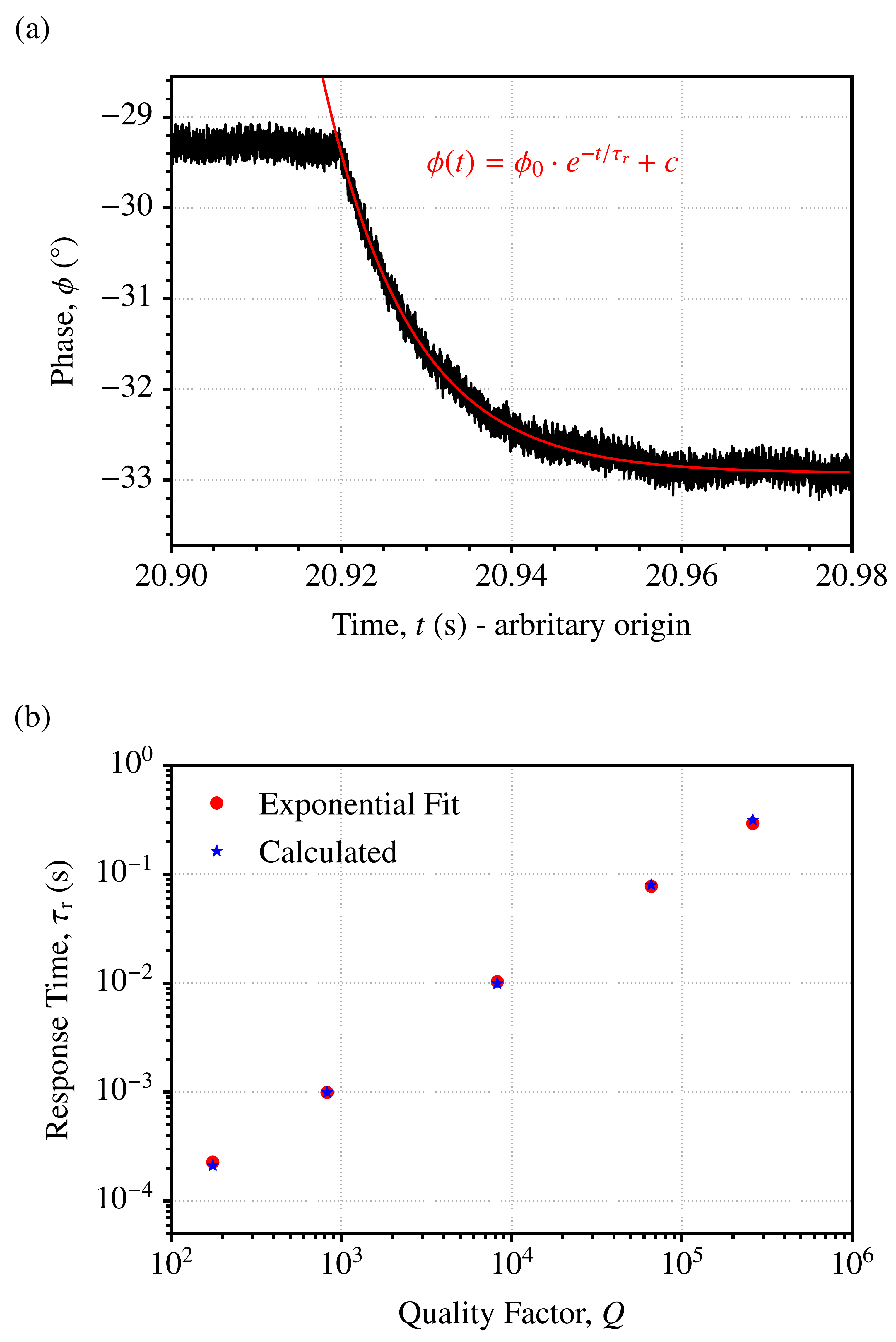}
  \caption[Figure Time]{Transient response of the phase. (a) Extraction of response times from an exponential fit to the phase change. (b) Dependence of the resonator time constant $\tau_\text{r}$ on the mechanical quality factor $Q$ as extracted from the exponential fits and calculations using $\tau_\text{r} = Q/\left(\pi f_\text{0}\right)$.}\label{fig:time}
\end{figure}

The $Q$-dependence of $\tau_\text{r}$ observed in Fig.~\ref{fig:Q}, indirectly through $\sigma_\text{A}$, can also be quantified directly through the response of the phase to external stimuli, be it mass, force, or temperature. In fact, not only the amplitude, but also the phase of a resonator responds to stimuli with a response time determined by the intrinsic resonator time constant $\tau_\text{r}$ \cite{Demir2020,Olcum2015}. This response can be measured experimentally in open-loop configuration by varying the piezo drive frequency and seeing how fast the phase responds, as is shown in Fig.~\ref{fig:time} for various $Q$-values. The response time was extracted by fitting an exponential decay to the phase data of the type $\phi\left(t\right) = \phi_\text{0} e^{-t/\tau_{\text{r}}} + c$. By varying the piezo drive frequency within the linear phase regime, a step response could be recorded with an example given in Fig.~\ref{fig:time}(a) along with an exponential fit. Fig.~\ref{fig:time}(b) shows measured $\tau_\text{r}$ versus $Q$ and the results are compared to values calculated using $\tau_\text{r} = Q/\left(\pi f_\text{0}\right)$. Good agreement can be observed, showing that the phase has a transient response in open-loop with a $Q$-dependent response time $\tau_\text{r}$.

\bibliography{Lib}

\begin{thebibliography}{35}%
\makeatletter
\providecommand \@ifxundefined [1]{%
 \@ifx{#1\undefined}
}%
\providecommand \@ifnum [1]{%
 \ifnum #1\expandafter \@firstoftwo
 \else \expandafter \@secondoftwo
 \fi
}%
\providecommand \@ifx [1]{%
 \ifx #1\expandafter \@firstoftwo
 \else \expandafter \@secondoftwo
 \fi
}%
\providecommand \natexlab [1]{#1}%
\providecommand \enquote  [1]{``#1''}%
\providecommand \bibnamefont  [1]{#1}%
\providecommand \bibfnamefont [1]{#1}%
\providecommand \citenamefont [1]{#1}%
\providecommand \href@noop [0]{\@secondoftwo}%
\providecommand \href [0]{\begingroup \@sanitize@url \@href}%
\providecommand \@href[1]{\@@startlink{#1}\@@href}%
\providecommand \@@href[1]{\endgroup#1\@@endlink}%
\providecommand \@sanitize@url [0]{\catcode `\\12\catcode `\$12\catcode
  `\&12\catcode `\#12\catcode `\^12\catcode `\_12\catcode `\%12\relax}%
\providecommand \@@startlink[1]{}%
\providecommand \@@endlink[0]{}%
\providecommand \url  [0]{\begingroup\@sanitize@url \@url }%
\providecommand \@url [1]{\endgroup\@href {#1}{\urlprefix }}%
\providecommand \urlprefix  [0]{URL }%
\providecommand \Eprint [0]{\href }%
\providecommand \doibase [0]{http://dx.doi.org/}%
\providecommand \selectlanguage [0]{\@gobble}%
\providecommand \bibinfo  [0]{\@secondoftwo}%
\providecommand \bibfield  [0]{\@secondoftwo}%
\providecommand \translation [1]{[#1]}%
\providecommand \BibitemOpen [0]{}%
\providecommand \bibitemStop [0]{}%
\providecommand \bibitemNoStop [0]{.\EOS\space}%
\providecommand \EOS [0]{\spacefactor3000\relax}%
\providecommand \BibitemShut  [1]{\csname bibitem#1\endcsname}%
\let\auto@bib@innerbib\@empty
\bibitem [{\citenamefont {Schmid}\ \emph {et~al.}(2016)\citenamefont {Schmid},
  \citenamefont {Villanueva},\ and\ \citenamefont {Roukes}}]{Schmid2016}%
  \BibitemOpen
  \bibfield  {author} {\bibinfo {author} {\bibfnamefont {S.}~\bibnamefont
  {Schmid}}, \bibinfo {author} {\bibfnamefont {L.~G.}\ \bibnamefont
  {Villanueva}}, \ and\ \bibinfo {author} {\bibfnamefont {M.~L.}\ \bibnamefont
  {Roukes}},\ }\href@noop {} {\emph {\bibinfo {title} {Fundamentals of
  nanomechanical resonators}}},\ \bibinfo {edition} {1st}\ ed.\ (\bibinfo
  {publisher} {Springer International Edition},\ \bibinfo {year}
  {2016})\BibitemShut {NoStop}%
\bibitem [{\citenamefont {Ekinci}\ and\ \citenamefont
  {Roukes}(2005)}]{Ekinci2005}%
  \BibitemOpen
  \bibfield  {author} {\bibinfo {author} {\bibfnamefont {K.~L.}\ \bibnamefont
  {Ekinci}}\ and\ \bibinfo {author} {\bibfnamefont {M.~L.}\ \bibnamefont
  {Roukes}},\ }\bibfield  {title} {\enquote {\bibinfo {title}
  {Nanoelectromechanical systems},}\ }\href {\doibase 10.1063/1.1927327}
  {\bibfield  {journal} {\bibinfo  {journal} {Review of Scientific
  Instruments}\ }\textbf {\bibinfo {volume} {76}},\ \bibinfo {pages} {061101}
  (\bibinfo {year} {2005})}\BibitemShut {NoStop}%
\bibitem [{\citenamefont {Ekinci}\ \emph
  {et~al.}(2004{\natexlab{a}})\citenamefont {Ekinci}, \citenamefont {Huang},\
  and\ \citenamefont {Roukes}}]{Ekinci2004_1}%
  \BibitemOpen
  \bibfield  {author} {\bibinfo {author} {\bibfnamefont {K.~L.}\ \bibnamefont
  {Ekinci}}, \bibinfo {author} {\bibfnamefont {X.~M.~H.}\ \bibnamefont
  {Huang}}, \ and\ \bibinfo {author} {\bibfnamefont {M.~L.}\ \bibnamefont
  {Roukes}},\ }\bibfield  {title} {\enquote {\bibinfo {title} {Ultrasensitive
  nanoelectromechanical mass detection},}\ }\href {\doibase 10.1063/1.1755417}
  {\bibfield  {journal} {\bibinfo  {journal} {Applied Physics Letters}\
  }\textbf {\bibinfo {volume} {84}},\ \bibinfo {pages} {4469--4471} (\bibinfo
  {year} {2004}{\natexlab{a}})}\BibitemShut {NoStop}%
\bibitem [{\citenamefont {Naik}\ \emph {et~al.}(2009)\citenamefont {Naik},
  \citenamefont {Hanay}, \citenamefont {Hiebert}, \citenamefont {Feng},\ and\
  \citenamefont {Roukes}}]{Naik2009}%
  \BibitemOpen
  \bibfield  {author} {\bibinfo {author} {\bibfnamefont {A.~K.}\ \bibnamefont
  {Naik}}, \bibinfo {author} {\bibfnamefont {M.~S.}\ \bibnamefont {Hanay}},
  \bibinfo {author} {\bibfnamefont {W.~K.}\ \bibnamefont {Hiebert}}, \bibinfo
  {author} {\bibfnamefont {X.~L.}\ \bibnamefont {Feng}}, \ and\ \bibinfo
  {author} {\bibfnamefont {M.~L.}\ \bibnamefont {Roukes}},\ }\bibfield  {title}
  {\enquote {\bibinfo {title} {Towards single-molecule nanomechanical mass
  spectrometry},}\ }\href {\doibase 10.1038/nnano.2009.152} {\bibfield
  {journal} {\bibinfo  {journal} {Nature Nanotechnology}\ }\textbf {\bibinfo
  {volume} {4}},\ \bibinfo {pages} {445--450} (\bibinfo {year}
  {2009})}\BibitemShut {NoStop}%
\bibitem [{\citenamefont {Chaste}\ \emph {et~al.}(2012)\citenamefont {Chaste},
  \citenamefont {Eichler}, \citenamefont {Moser}, \citenamefont {Ceballos},
  \citenamefont {Rurali},\ and\ \citenamefont {Bachtold}}]{Chaste2012}%
  \BibitemOpen
  \bibfield  {author} {\bibinfo {author} {\bibfnamefont {J.}~\bibnamefont
  {Chaste}}, \bibinfo {author} {\bibfnamefont {A.}~\bibnamefont {Eichler}},
  \bibinfo {author} {\bibfnamefont {J.}~\bibnamefont {Moser}}, \bibinfo
  {author} {\bibfnamefont {G.}~\bibnamefont {Ceballos}}, \bibinfo {author}
  {\bibfnamefont {R.}~\bibnamefont {Rurali}}, \ and\ \bibinfo {author}
  {\bibfnamefont {A.}~\bibnamefont {Bachtold}},\ }\bibfield  {title} {\enquote
  {\bibinfo {title} {A nanomechanical mass sensor with yoctogram resolution},}\
  }\href {\doibase 10.1038/nnano.2012.42} {\bibfield  {journal} {\bibinfo
  {journal} {Nature Nanotechnology}\ }\textbf {\bibinfo {volume} {7}},\
  \bibinfo {pages} {301--304} (\bibinfo {year} {2012})}\BibitemShut {NoStop}%
\bibitem [{\citenamefont {Kozinsky}\ \emph {et~al.}(2006)\citenamefont
  {Kozinsky}, \citenamefont {Postma}, \citenamefont {Bargatin},\ and\
  \citenamefont {Roukes}}]{Kozinsky2006}%
  \BibitemOpen
  \bibfield  {author} {\bibinfo {author} {\bibfnamefont {I.}~\bibnamefont
  {Kozinsky}}, \bibinfo {author} {\bibfnamefont {H.~W.~C.}\ \bibnamefont
  {Postma}}, \bibinfo {author} {\bibfnamefont {I.}~\bibnamefont {Bargatin}}, \
  and\ \bibinfo {author} {\bibfnamefont {M.~L.}\ \bibnamefont {Roukes}},\
  }\bibfield  {title} {\enquote {\bibinfo {title} {Tuning nonlinearity, dynamic
  range, and frequency of nanomechanical resonators},}\ }\href {\doibase
  10.1063/1.2209211} {\bibfield  {journal} {\bibinfo  {journal} {Applied
  Physics Letters}\ }\textbf {\bibinfo {volume} {88}},\ \bibinfo {pages}
  {253101} (\bibinfo {year} {2006})}\BibitemShut {NoStop}%
\bibitem [{\citenamefont {Moser}\ \emph {et~al.}(2013)\citenamefont {Moser},
  \citenamefont {G\"{u}ttinger}, \citenamefont {Eichler}, \citenamefont
  {Esplandiu}, \citenamefont {Liu}, \citenamefont {Dykman},\ and\ \citenamefont
  {Bachtold}}]{Moser2013}%
  \BibitemOpen
  \bibfield  {author} {\bibinfo {author} {\bibfnamefont {J.}~\bibnamefont
  {Moser}}, \bibinfo {author} {\bibfnamefont {J.}~\bibnamefont
  {G\"{u}ttinger}}, \bibinfo {author} {\bibfnamefont {A.}~\bibnamefont
  {Eichler}}, \bibinfo {author} {\bibfnamefont {M.~J.}\ \bibnamefont
  {Esplandiu}}, \bibinfo {author} {\bibfnamefont {D.~E.}\ \bibnamefont {Liu}},
  \bibinfo {author} {\bibfnamefont {M.~I.}\ \bibnamefont {Dykman}}, \ and\
  \bibinfo {author} {\bibfnamefont {A.}~\bibnamefont {Bachtold}},\ }\bibfield
  {title} {\enquote {\bibinfo {title} {Ultrasensitive force detection with a
  nanotube mechanical resonator},}\ }\href {\doibase 10.1038/nnano.2013.97}
  {\bibfield  {journal} {\bibinfo  {journal} {Nature Nanotechnology}\ }\textbf
  {\bibinfo {volume} {8}},\ \bibinfo {pages} {493--496} (\bibinfo {year}
  {2013})}\BibitemShut {NoStop}%
\bibitem [{\citenamefont {Pandey}\ \emph {et~al.}(2010)\citenamefont {Pandey},
  \citenamefont {Gottlieb}, \citenamefont {Shtempluck},\ and\ \citenamefont
  {Buks}}]{Pandey2010}%
  \BibitemOpen
  \bibfield  {author} {\bibinfo {author} {\bibfnamefont {A.~K.}\ \bibnamefont
  {Pandey}}, \bibinfo {author} {\bibfnamefont {O.}~\bibnamefont {Gottlieb}},
  \bibinfo {author} {\bibfnamefont {O.}~\bibnamefont {Shtempluck}}, \ and\
  \bibinfo {author} {\bibfnamefont {E.}~\bibnamefont {Buks}},\ }\bibfield
  {title} {\enquote {\bibinfo {title} {Performance of an {A}u{P}d
  micromechanical resonator as a temperature sensor},}\ }\href {\doibase
  10.1063/1.3431614} {\bibfield  {journal} {\bibinfo  {journal} {Applied
  Physics Letters}\ }\textbf {\bibinfo {volume} {96}},\ \bibinfo {pages}
  {203105} (\bibinfo {year} {2010})}\BibitemShut {NoStop}%
\bibitem [{\citenamefont {Larsen}\ \emph {et~al.}(2011)\citenamefont {Larsen},
  \citenamefont {Schmid}, \citenamefont {Grönberg}, \citenamefont {Niskanen},
  \citenamefont {Hassel}, \citenamefont {Dohn},\ and\ \citenamefont
  {Boisen}}]{Larsen2011}%
  \BibitemOpen
  \bibfield  {author} {\bibinfo {author} {\bibfnamefont {T.}~\bibnamefont
  {Larsen}}, \bibinfo {author} {\bibfnamefont {S.}~\bibnamefont {Schmid}},
  \bibinfo {author} {\bibfnamefont {L.}~\bibnamefont {Grönberg}}, \bibinfo
  {author} {\bibfnamefont {A.~O.}\ \bibnamefont {Niskanen}}, \bibinfo {author}
  {\bibfnamefont {J.}~\bibnamefont {Hassel}}, \bibinfo {author} {\bibfnamefont
  {S.}~\bibnamefont {Dohn}}, \ and\ \bibinfo {author} {\bibfnamefont
  {A.}~\bibnamefont {Boisen}},\ }\bibfield  {title} {\enquote {\bibinfo {title}
  {Ultrasensitive string-based temperature sensors},}\ }\href {\doibase
  10.1063/1.3567012} {\bibfield  {journal} {\bibinfo  {journal} {Applied
  Physics Letters}\ }\textbf {\bibinfo {volume} {98}},\ \bibinfo {pages}
  {121901} (\bibinfo {year} {2011})}\BibitemShut {NoStop}%
\bibitem [{\citenamefont {Zhang}\ \emph {et~al.}(2013)\citenamefont {Zhang},
  \citenamefont {Myers}, \citenamefont {Sader},\ and\ \citenamefont
  {Roukes}}]{Zhang2013}%
  \BibitemOpen
  \bibfield  {author} {\bibinfo {author} {\bibfnamefont {X.~C.}\ \bibnamefont
  {Zhang}}, \bibinfo {author} {\bibfnamefont {E.~B.}\ \bibnamefont {Myers}},
  \bibinfo {author} {\bibfnamefont {J.~E.}\ \bibnamefont {Sader}}, \ and\
  \bibinfo {author} {\bibfnamefont {M.~L.}\ \bibnamefont {Roukes}},\ }\bibfield
   {title} {\enquote {\bibinfo {title} {Nanomechanical torsional resonators for
  frequency-shift infrared thermal sensing},}\ }\href {\doibase
  10.1021/nl304687p} {\bibfield  {journal} {\bibinfo  {journal} {Nano Letters}\
  }\textbf {\bibinfo {volume} {13}},\ \bibinfo {pages} {1528--1534} (\bibinfo
  {year} {2013})}\BibitemShut {NoStop}%
\bibitem [{\citenamefont {Piller}\ \emph {et~al.}(2019)\citenamefont {Piller},
  \citenamefont {Luhmann}, \citenamefont {Chien},\ and\ \citenamefont
  {Schmid}}]{Piller2019}%
  \BibitemOpen
  \bibfield  {author} {\bibinfo {author} {\bibfnamefont {M.}~\bibnamefont
  {Piller}}, \bibinfo {author} {\bibfnamefont {N.}~\bibnamefont {Luhmann}},
  \bibinfo {author} {\bibfnamefont {M.-H.}\ \bibnamefont {Chien}}, \ and\
  \bibinfo {author} {\bibfnamefont {S.}~\bibnamefont {Schmid}},\ }\bibfield
  {title} {\enquote {\bibinfo {title} {{Nanoelectromechanical infrared
  detector}},}\ }in\ \href {\doibase 10.1117/12.2528416} {\emph {\bibinfo
  {booktitle} {Optical Sensing, Imaging, and Photon Counting: From X-Rays to
  THz 2019}}},\ Vol.\ \bibinfo {volume} {11088},\ \bibinfo {editor} {edited by\
  \bibinfo {editor} {\bibfnamefont {O.}~\bibnamefont {Mitrofanov}}},\ \bibinfo
  {organization} {International Society for Optics and Photonics}\ (\bibinfo
  {publisher} {SPIE},\ \bibinfo {year} {2019})\ pp.\ \bibinfo {pages} {9 --
  15}\BibitemShut {NoStop}%
\bibitem [{\citenamefont {Ekinci}\ \emph
  {et~al.}(2004{\natexlab{b}})\citenamefont {Ekinci}, \citenamefont {Yang},\
  and\ \citenamefont {Roukes}}]{Ekinci2004_2}%
  \BibitemOpen
  \bibfield  {author} {\bibinfo {author} {\bibfnamefont {K.~L.}\ \bibnamefont
  {Ekinci}}, \bibinfo {author} {\bibfnamefont {Y.~T.}\ \bibnamefont {Yang}}, \
  and\ \bibinfo {author} {\bibfnamefont {M.~L.}\ \bibnamefont {Roukes}},\
  }\bibfield  {title} {\enquote {\bibinfo {title} {Ultimate limits to inertial
  mass sensing based upon nanoelectromechanical systems},}\ }\href {\doibase
  10.1063/1.1642738} {\bibfield  {journal} {\bibinfo  {journal} {Journal of
  Applied Physics}\ }\textbf {\bibinfo {volume} {95}},\ \bibinfo {pages}
  {2682--2689} (\bibinfo {year} {2004}{\natexlab{b}})}\BibitemShut {NoStop}%
\bibitem [{\citenamefont {{Vig}}\ and\ \citenamefont {{Yoonkee
  Kim}}(1999)}]{Vig1999}%
  \BibitemOpen
  \bibfield  {author} {\bibinfo {author} {\bibfnamefont {J.~R.}\ \bibnamefont
  {{Vig}}}\ and\ \bibinfo {author} {\bibnamefont {{Yoonkee Kim}}},\ }\bibfield
  {title} {\enquote {\bibinfo {title} {Noise in microelectromechanical system
  resonators},}\ }\href@noop {} {\bibfield  {journal} {\bibinfo  {journal}
  {IEEE Transactions on Ultrasonics, Ferroelectrics, and Frequency Control}\
  }\textbf {\bibinfo {volume} {46}},\ \bibinfo {pages} {1558--1565} (\bibinfo
  {year} {1999})}\BibitemShut {NoStop}%
\bibitem [{\citenamefont {Cleland}\ and\ \citenamefont
  {Roukes}(2002)}]{Cleland2002}%
  \BibitemOpen
  \bibfield  {author} {\bibinfo {author} {\bibfnamefont {A.~N.}\ \bibnamefont
  {Cleland}}\ and\ \bibinfo {author} {\bibfnamefont {M.~L.}\ \bibnamefont
  {Roukes}},\ }\bibfield  {title} {\enquote {\bibinfo {title} {Noise processes
  in nanomechanical resonators},}\ }\href {\doibase 10.1063/1.1499745}
  {\bibfield  {journal} {\bibinfo  {journal} {Journal of Applied Physics}\
  }\textbf {\bibinfo {volume} {92}},\ \bibinfo {pages} {2758--2769} (\bibinfo
  {year} {2002})}\BibitemShut {NoStop}%
\bibitem [{\citenamefont {Atalaya}\ \emph {et~al.}(2011)\citenamefont
  {Atalaya}, \citenamefont {Isacsson},\ and\ \citenamefont
  {Dykman}}]{Atalaya2011}%
  \BibitemOpen
  \bibfield  {author} {\bibinfo {author} {\bibfnamefont {J.}~\bibnamefont
  {Atalaya}}, \bibinfo {author} {\bibfnamefont {A.}~\bibnamefont {Isacsson}}, \
  and\ \bibinfo {author} {\bibfnamefont {M.~I.}\ \bibnamefont {Dykman}},\
  }\bibfield  {title} {\enquote {\bibinfo {title} {Diffusion-induced dephasing
  in nanomechanical resonators},}\ }\href {\doibase 10.1103/PhysRevB.83.045419}
  {\bibfield  {journal} {\bibinfo  {journal} {Physical Review B}\ }\textbf
  {\bibinfo {volume} {83}},\ \bibinfo {pages} {045419} (\bibinfo {year}
  {2011})}\BibitemShut {NoStop}%
\bibitem [{\citenamefont {Maillet}\ \emph {et~al.}(2018)\citenamefont
  {Maillet}, \citenamefont {Zhou}, \citenamefont {Gazizulin}, \citenamefont
  {Ilic}, \citenamefont {Parpia}, \citenamefont {Bourgeois}, \citenamefont
  {Fefferman},\ and\ \citenamefont {Collin}}]{Maillet2018}%
  \BibitemOpen
  \bibfield  {author} {\bibinfo {author} {\bibfnamefont {O.}~\bibnamefont
  {Maillet}}, \bibinfo {author} {\bibfnamefont {X.}~\bibnamefont {Zhou}},
  \bibinfo {author} {\bibfnamefont {R.~R.}\ \bibnamefont {Gazizulin}}, \bibinfo
  {author} {\bibfnamefont {R.}~\bibnamefont {Ilic}}, \bibinfo {author}
  {\bibfnamefont {J.~M.}\ \bibnamefont {Parpia}}, \bibinfo {author}
  {\bibfnamefont {O.}~\bibnamefont {Bourgeois}}, \bibinfo {author}
  {\bibfnamefont {A.~D.}\ \bibnamefont {Fefferman}}, \ and\ \bibinfo {author}
  {\bibfnamefont {E.}~\bibnamefont {Collin}},\ }\bibfield  {title} {\enquote
  {\bibinfo {title} {Measuring frequency fluctuations in nonlinear
  nanomechanical resonators},}\ }\href {\doibase 10.1021/acsnano.8b01634}
  {\bibfield  {journal} {\bibinfo  {journal} {{ACS} Nano}\ }\textbf {\bibinfo
  {volume} {12}},\ \bibinfo {pages} {5753--5760} (\bibinfo {year}
  {2018})}\BibitemShut {NoStop}%
\bibitem [{\citenamefont {Robins}(1984)}]{Robins1984}%
  \BibitemOpen
  \bibfield  {author} {\bibinfo {author} {\bibfnamefont {W.}~\bibnamefont
  {Robins}},\ }\href@noop {} {\emph {\bibinfo {title} {Phase noise in signal
  sources: theory and applications}}},\ Vol.~\bibinfo {volume} {9}\ (\bibinfo
  {publisher} {IET},\ \bibinfo {year} {1984})\BibitemShut {NoStop}%
\bibitem [{\citenamefont {Albrecht}\ \emph {et~al.}(1991)\citenamefont
  {Albrecht}, \citenamefont {Grütter}, \citenamefont {Horne},\ and\
  \citenamefont {Rugar}}]{Albrecht1991}%
  \BibitemOpen
  \bibfield  {author} {\bibinfo {author} {\bibfnamefont {T.~R.}\ \bibnamefont
  {Albrecht}}, \bibinfo {author} {\bibfnamefont {P.}~\bibnamefont {Grütter}},
  \bibinfo {author} {\bibfnamefont {D.}~\bibnamefont {Horne}}, \ and\ \bibinfo
  {author} {\bibfnamefont {D.}~\bibnamefont {Rugar}},\ }\bibfield  {title}
  {\enquote {\bibinfo {title} {Frequency modulation detection using
  high‐{$Q$} cantilevers for enhanced force microscope sensitivity},}\ }\href
  {\doibase 10.1063/1.347347} {\bibfield  {journal} {\bibinfo  {journal}
  {Journal of Applied Physics}\ }\textbf {\bibinfo {volume} {69}},\ \bibinfo
  {pages} {668--673} (\bibinfo {year} {1991})}\BibitemShut {NoStop}%
\bibitem [{\citenamefont {Cleland}(2005)}]{Cleland2005}%
  \BibitemOpen
  \bibfield  {author} {\bibinfo {author} {\bibfnamefont {A.~N.}\ \bibnamefont
  {Cleland}},\ }\bibfield  {title} {\enquote {\bibinfo {title}
  {Thermomechanical noise limits on parametric sensing with nanomechanical
  resonators},}\ }\href {\doibase 10.1088/1367-2630/7/1/235} {\bibfield
  {journal} {\bibinfo  {journal} {New Journal of Physics}\ }\textbf {\bibinfo
  {volume} {7}},\ \bibinfo {pages} {235--235} (\bibinfo {year}
  {2005})}\BibitemShut {NoStop}%
\bibitem [{\citenamefont {Gavartin}\ \emph {et~al.}(2013)\citenamefont
  {Gavartin}, \citenamefont {Verlot},\ and\ \citenamefont
  {Kippenberg}}]{Gavartin2013}%
  \BibitemOpen
  \bibfield  {author} {\bibinfo {author} {\bibfnamefont {E.}~\bibnamefont
  {Gavartin}}, \bibinfo {author} {\bibfnamefont {P.}~\bibnamefont {Verlot}}, \
  and\ \bibinfo {author} {\bibfnamefont {T.~J.}\ \bibnamefont {Kippenberg}},\
  }\bibfield  {title} {\enquote {\bibinfo {title} {Stabilization of a linear
  nanomechanical oscillator to its thermodynamic limit},}\ }\href {\doibase
  10.1038/ncomms3860} {\bibfield  {journal} {\bibinfo  {journal} {Nature
  Communications}\ }\textbf {\bibinfo {volume} {4}},\ \bibinfo {pages} {2860}
  (\bibinfo {year} {2013})}\BibitemShut {NoStop}%
\bibitem [{\citenamefont {Sansa}\ \emph {et~al.}(2016)\citenamefont {Sansa},
  \citenamefont {Sage}, \citenamefont {Bullard}, \citenamefont {G{\'{e}}ly},
  \citenamefont {Alava}, \citenamefont {Colinet}, \citenamefont {Naik},
  \citenamefont {Villanueva}, \citenamefont {Duraffourg}, \citenamefont
  {Roukes}, \citenamefont {Jourdan},\ and\ \citenamefont {Hentz}}]{Sansa2016}%
  \BibitemOpen
  \bibfield  {author} {\bibinfo {author} {\bibfnamefont {M.}~\bibnamefont
  {Sansa}}, \bibinfo {author} {\bibfnamefont {E.}~\bibnamefont {Sage}},
  \bibinfo {author} {\bibfnamefont {E.~C.}\ \bibnamefont {Bullard}}, \bibinfo
  {author} {\bibfnamefont {M.}~\bibnamefont {G{\'{e}}ly}}, \bibinfo {author}
  {\bibfnamefont {T.}~\bibnamefont {Alava}}, \bibinfo {author} {\bibfnamefont
  {E.}~\bibnamefont {Colinet}}, \bibinfo {author} {\bibfnamefont {A.~K.}\
  \bibnamefont {Naik}}, \bibinfo {author} {\bibfnamefont {L.~G.}\ \bibnamefont
  {Villanueva}}, \bibinfo {author} {\bibfnamefont {L.}~\bibnamefont
  {Duraffourg}}, \bibinfo {author} {\bibfnamefont {M.~L.}\ \bibnamefont
  {Roukes}}, \bibinfo {author} {\bibfnamefont {G.}~\bibnamefont {Jourdan}}, \
  and\ \bibinfo {author} {\bibfnamefont {S.}~\bibnamefont {Hentz}},\ }\bibfield
   {title} {\enquote {\bibinfo {title} {Frequency fluctuations in silicon
  nanoresonators},}\ }\href {\doibase 10.1038/nnano.2016.19} {\bibfield
  {journal} {\bibinfo  {journal} {Nature Nanotechnology}\ }\textbf {\bibinfo
  {volume} {11}},\ \bibinfo {pages} {552--558} (\bibinfo {year}
  {2016})}\BibitemShut {NoStop}%
\bibitem [{\citenamefont {Tsaturyan}\ \emph {et~al.}(2017)\citenamefont
  {Tsaturyan}, \citenamefont {Barg}, \citenamefont {Polzik},\ and\
  \citenamefont {Schliesser}}]{Tsaturyan2017}%
  \BibitemOpen
  \bibfield  {author} {\bibinfo {author} {\bibfnamefont {Y.}~\bibnamefont
  {Tsaturyan}}, \bibinfo {author} {\bibfnamefont {A.}~\bibnamefont {Barg}},
  \bibinfo {author} {\bibfnamefont {E.~S.}\ \bibnamefont {Polzik}}, \ and\
  \bibinfo {author} {\bibfnamefont {A.}~\bibnamefont {Schliesser}},\ }\bibfield
   {title} {\enquote {\bibinfo {title} {Ultracoherent nanomechanical resonators
  via soft clamping and dissipation dilution},}\ }\href {\doibase
  10.1038/nnano.2017.101} {\bibfield  {journal} {\bibinfo  {journal} {Nature
  Nanotechnology}\ }\textbf {\bibinfo {volume} {12}},\ \bibinfo {pages}
  {776--783} (\bibinfo {year} {2017})}\BibitemShut {NoStop}%
\bibitem [{\citenamefont {Ghadimi}\ \emph {et~al.}(2018)\citenamefont
  {Ghadimi}, \citenamefont {Fedorov}, \citenamefont {Engelsen}, \citenamefont
  {Bereyhi}, \citenamefont {Schilling}, \citenamefont {Wilson},\ and\
  \citenamefont {Kippenberg}}]{Ghadimi2018}%
  \BibitemOpen
  \bibfield  {author} {\bibinfo {author} {\bibfnamefont {A.~H.}\ \bibnamefont
  {Ghadimi}}, \bibinfo {author} {\bibfnamefont {S.~A.}\ \bibnamefont
  {Fedorov}}, \bibinfo {author} {\bibfnamefont {N.~J.}\ \bibnamefont
  {Engelsen}}, \bibinfo {author} {\bibfnamefont {M.~J.}\ \bibnamefont
  {Bereyhi}}, \bibinfo {author} {\bibfnamefont {R.}~\bibnamefont {Schilling}},
  \bibinfo {author} {\bibfnamefont {D.~J.}\ \bibnamefont {Wilson}}, \ and\
  \bibinfo {author} {\bibfnamefont {T.~J.}\ \bibnamefont {Kippenberg}},\
  }\bibfield  {title} {\enquote {\bibinfo {title} {Elastic strain engineering
  for ultralow mechanical dissipation},}\ }\href {\doibase
  10.1126/science.aar6939} {\bibfield  {journal} {\bibinfo  {journal}
  {Science}\ }\textbf {\bibinfo {volume} {360}},\ \bibinfo {pages} {764--768}
  (\bibinfo {year} {2018})}\BibitemShut {NoStop}%
\bibitem [{\citenamefont {Fong}\ \emph {et~al.}(2012)\citenamefont {Fong},
  \citenamefont {Pernice},\ and\ \citenamefont {Tang}}]{Fong2012}%
  \BibitemOpen
  \bibfield  {author} {\bibinfo {author} {\bibfnamefont {K.~Y.}\ \bibnamefont
  {Fong}}, \bibinfo {author} {\bibfnamefont {W.~H.~P.}\ \bibnamefont
  {Pernice}}, \ and\ \bibinfo {author} {\bibfnamefont {H.~X.}\ \bibnamefont
  {Tang}},\ }\bibfield  {title} {\enquote {\bibinfo {title} {Frequency and
  phase noise of ultrahigh {$Q$} silicon nitride nanomechanical resonators},}\
  }\href {\doibase 10.1103/PhysRevB.85.161410} {\bibfield  {journal} {\bibinfo
  {journal} {Physical Review B}\ }\textbf {\bibinfo {volume} {85}},\ \bibinfo
  {pages} {161410(R)} (\bibinfo {year} {2012})}\BibitemShut {NoStop}%
\bibitem [{\citenamefont {Roy}\ \emph {et~al.}(2018)\citenamefont {Roy},
  \citenamefont {Sauer}, \citenamefont {Westwood-Bachman}, \citenamefont
  {Venkatasubramanian},\ and\ \citenamefont {Hiebert}}]{Roy2018}%
  \BibitemOpen
  \bibfield  {author} {\bibinfo {author} {\bibfnamefont {S.~K.}\ \bibnamefont
  {Roy}}, \bibinfo {author} {\bibfnamefont {V.~T.~K.}\ \bibnamefont {Sauer}},
  \bibinfo {author} {\bibfnamefont {J.~N.}\ \bibnamefont {Westwood-Bachman}},
  \bibinfo {author} {\bibfnamefont {A.}~\bibnamefont {Venkatasubramanian}}, \
  and\ \bibinfo {author} {\bibfnamefont {W.~K.}\ \bibnamefont {Hiebert}},\
  }\bibfield  {title} {\enquote {\bibinfo {title} {Improving mechanical sensor
  performance through larger damping},}\ }\href {\doibase
  10.1126/science.aar5220} {\bibfield  {journal} {\bibinfo  {journal}
  {Science}\ }\textbf {\bibinfo {volume} {360}},\ \bibinfo {pages} {eaar5220}
  (\bibinfo {year} {2018})}\BibitemShut {NoStop}%
\bibitem [{\citenamefont {{Demir}}\ and\ \citenamefont
  {{Hanay}}(2020)}]{Demir2020}%
  \BibitemOpen
  \bibfield  {author} {\bibinfo {author} {\bibfnamefont {A.}~\bibnamefont
  {{Demir}}}\ and\ \bibinfo {author} {\bibfnamefont {M.~S.}\ \bibnamefont
  {{Hanay}}},\ }\bibfield  {title} {\enquote {\bibinfo {title} {Fundamental
  sensitivity limitations of nanomechanical resonant sensors due to
  thermomechanical noise},}\ }\href
  {https://ieeexplore.ieee.org/document/8878147} {\bibfield  {journal}
  {\bibinfo  {journal} {IEEE Sensors Journal}\ }\textbf {\bibinfo {volume}
  {20}},\ \bibinfo {pages} {1947--1961} (\bibinfo {year} {2020})}\BibitemShut
  {NoStop}%
\bibitem [{\citenamefont {Verbridge}\ \emph {et~al.}(2006)\citenamefont
  {Verbridge}, \citenamefont {Parpia}, \citenamefont {Reichenbach},
  \citenamefont {Bellan},\ and\ \citenamefont {Craighead}}]{Verbridge2006}%
  \BibitemOpen
  \bibfield  {author} {\bibinfo {author} {\bibfnamefont {S.~S.}\ \bibnamefont
  {Verbridge}}, \bibinfo {author} {\bibfnamefont {J.~M.}\ \bibnamefont
  {Parpia}}, \bibinfo {author} {\bibfnamefont {R.~B.}\ \bibnamefont
  {Reichenbach}}, \bibinfo {author} {\bibfnamefont {L.~M.}\ \bibnamefont
  {Bellan}}, \ and\ \bibinfo {author} {\bibfnamefont {H.~G.}\ \bibnamefont
  {Craighead}},\ }\bibfield  {title} {\enquote {\bibinfo {title} {High quality
  factor resonance at room temperature with nanostrings under high tensile
  stress},}\ }\href {\doibase 10.1063/1.2204829} {\bibfield  {journal}
  {\bibinfo  {journal} {Journal of Applied Physics}\ }\textbf {\bibinfo
  {volume} {99}},\ \bibinfo {pages} {124304} (\bibinfo {year}
  {2006})}\BibitemShut {NoStop}%
\bibitem [{\citenamefont {Schmid}\ \emph {et~al.}(2011)\citenamefont {Schmid},
  \citenamefont {Jensen}, \citenamefont {Nielsen},\ and\ \citenamefont
  {Boisen}}]{Schmid2011}%
  \BibitemOpen
  \bibfield  {author} {\bibinfo {author} {\bibfnamefont {S.}~\bibnamefont
  {Schmid}}, \bibinfo {author} {\bibfnamefont {K.~D.}\ \bibnamefont {Jensen}},
  \bibinfo {author} {\bibfnamefont {K.~H.}\ \bibnamefont {Nielsen}}, \ and\
  \bibinfo {author} {\bibfnamefont {A.}~\bibnamefont {Boisen}},\ }\bibfield
  {title} {\enquote {\bibinfo {title} {Damping mechanisms in high-{$Q$} micro
  and nanomechanical string resonators},}\ }\href {\doibase
  10.1103/PhysRevB.84.165307} {\bibfield  {journal} {\bibinfo  {journal}
  {Physical Review B}\ }\textbf {\bibinfo {volume} {84}},\ \bibinfo {pages}
  {165307} (\bibinfo {year} {2011})}\BibitemShut {NoStop}%
\bibitem [{\citenamefont {Unterreithmeier}\ \emph {et~al.}(2010)\citenamefont
  {Unterreithmeier}, \citenamefont {Faust},\ and\ \citenamefont
  {Kotthaus}}]{Unterreithmeier2010}%
  \BibitemOpen
  \bibfield  {author} {\bibinfo {author} {\bibfnamefont {Q.~P.}\ \bibnamefont
  {Unterreithmeier}}, \bibinfo {author} {\bibfnamefont {T.}~\bibnamefont
  {Faust}}, \ and\ \bibinfo {author} {\bibfnamefont {J.~P.}\ \bibnamefont
  {Kotthaus}},\ }\bibfield  {title} {\enquote {\bibinfo {title} {Damping of
  nanomechanical resonators},}\ }\href {\doibase
  10.1103/PhysRevLett.105.027205} {\bibfield  {journal} {\bibinfo  {journal}
  {Physical Review Letters}\ }\textbf {\bibinfo {volume} {105}},\ \bibinfo
  {pages} {027205} (\bibinfo {year} {2010})}\BibitemShut {NoStop}%
\bibitem [{\citenamefont {Schmid}\ \emph {et~al.}(2010)\citenamefont {Schmid},
  \citenamefont {Dohn},\ and\ \citenamefont {Boisen}}]{Schmid2010}%
  \BibitemOpen
  \bibfield  {author} {\bibinfo {author} {\bibfnamefont {S.}~\bibnamefont
  {Schmid}}, \bibinfo {author} {\bibfnamefont {S.}~\bibnamefont {Dohn}}, \ and\
  \bibinfo {author} {\bibfnamefont {A.}~\bibnamefont {Boisen}},\ }\bibfield
  {title} {\enquote {\bibinfo {title} {Real-time particle mass spectrometry
  based on resonant micro strings},}\ }\href {\doibase 10.3390/s100908092}
  {\bibfield  {journal} {\bibinfo  {journal} {Sensors}\ }\textbf {\bibinfo
  {volume} {10}},\ \bibinfo {pages} {8092–8100} (\bibinfo {year}
  {2010})}\BibitemShut {NoStop}%
\bibitem [{\citenamefont {{Allan}}(1966)}]{Allan1966}%
  \BibitemOpen
  \bibfield  {author} {\bibinfo {author} {\bibfnamefont {D.~W.}\ \bibnamefont
  {{Allan}}},\ }\bibfield  {title} {\enquote {\bibinfo {title} {Statistics of
  atomic frequency standards},}\ }\href@noop {} {\bibfield  {journal} {\bibinfo
   {journal} {Proceedings of the IEEE}\ }\textbf {\bibinfo {volume} {54}},\
  \bibinfo {pages} {221--230} (\bibinfo {year} {1966})}\BibitemShut {NoStop}%
\bibitem [{\citenamefont {Larsen}\ \emph {et~al.}(2013)\citenamefont {Larsen},
  \citenamefont {Schmid}, \citenamefont {Villanueva},\ and\ \citenamefont
  {Boisen}}]{Larsen2013}%
  \BibitemOpen
  \bibfield  {author} {\bibinfo {author} {\bibfnamefont {T.}~\bibnamefont
  {Larsen}}, \bibinfo {author} {\bibfnamefont {S.}~\bibnamefont {Schmid}},
  \bibinfo {author} {\bibfnamefont {L.~G.}\ \bibnamefont {Villanueva}}, \ and\
  \bibinfo {author} {\bibfnamefont {A.}~\bibnamefont {Boisen}},\ }\bibfield
  {title} {\enquote {\bibinfo {title} {{Photothermal analysis of individual
  nanoparticulate samples using micromechanical resonators.}}}\ }\href
  {\doibase 10.1021/nn402057f} {\bibfield  {journal} {\bibinfo  {journal} {ACS
  Nano}\ }\textbf {\bibinfo {volume} {7}},\ \bibinfo {pages} {6188--6193}
  (\bibinfo {year} {2013})}\BibitemShut {NoStop}%
\bibitem [{\citenamefont {Piller}\ \emph {et~al.}(2020)\citenamefont {Piller},
  \citenamefont {Sadeghi}, \citenamefont {West}, \citenamefont {Luhmann},
  \citenamefont {Martini}, \citenamefont {Hansen},\ and\ \citenamefont
  {Schmid}}]{Piller2020}%
  \BibitemOpen
  \bibfield  {author} {\bibinfo {author} {\bibfnamefont {M.}~\bibnamefont
  {Piller}}, \bibinfo {author} {\bibfnamefont {P.}~\bibnamefont {Sadeghi}},
  \bibinfo {author} {\bibfnamefont {R.~G.}\ \bibnamefont {West}}, \bibinfo
  {author} {\bibfnamefont {N.}~\bibnamefont {Luhmann}}, \bibinfo {author}
  {\bibfnamefont {P.}~\bibnamefont {Martini}}, \bibinfo {author} {\bibfnamefont
  {O.}~\bibnamefont {Hansen}}, \ and\ \bibinfo {author} {\bibfnamefont
  {S.}~\bibnamefont {Schmid}},\ }\bibfield  {title} {\enquote {\bibinfo {title}
  {Thermal radiation dominated heat transfer in nanomechanical silicon nitride
  drum resonators},}\ }\href {\doibase 10.1063/5.0015166} {\bibfield  {journal}
  {\bibinfo  {journal} {Applied Physics Letters}\ }\textbf {\bibinfo {volume}
  {117}},\ \bibinfo {pages} {034101} (\bibinfo {year} {2020})}\BibitemShut
  {NoStop}%
\bibitem [{\citenamefont {Verbridge}\ \emph {et~al.}(2008)\citenamefont
  {Verbridge}, \citenamefont {Ilic}, \citenamefont {Craighead},\ and\
  \citenamefont {Parpia}}]{Verbridge2008_2}%
  \BibitemOpen
  \bibfield  {author} {\bibinfo {author} {\bibfnamefont {S.~S.}\ \bibnamefont
  {Verbridge}}, \bibinfo {author} {\bibfnamefont {R.}~\bibnamefont {Ilic}},
  \bibinfo {author} {\bibfnamefont {H.~G.}\ \bibnamefont {Craighead}}, \ and\
  \bibinfo {author} {\bibfnamefont {J.~M.}\ \bibnamefont {Parpia}},\ }\bibfield
   {title} {\enquote {\bibinfo {title} {Size and frequency dependent gas
  damping of nanomechanical resonators},}\ }\href {\doibase 10.1063/1.2952762}
  {\bibfield  {journal} {\bibinfo  {journal} {Applied Physics Letters}\
  }\textbf {\bibinfo {volume} {93}},\ \bibinfo {pages} {013101} (\bibinfo
  {year} {2008})}\BibitemShut {NoStop}%
\bibitem [{\citenamefont {Olcum}\ \emph {et~al.}(2015)\citenamefont {Olcum},
  \citenamefont {Cermak}, \citenamefont {Wasserman},\ and\ \citenamefont
  {Manalis}}]{Olcum2015}%
  \BibitemOpen
  \bibfield  {author} {\bibinfo {author} {\bibfnamefont {S.}~\bibnamefont
  {Olcum}}, \bibinfo {author} {\bibfnamefont {N.}~\bibnamefont {Cermak}},
  \bibinfo {author} {\bibfnamefont {S.~C.}\ \bibnamefont {Wasserman}}, \ and\
  \bibinfo {author} {\bibfnamefont {S.~R.}\ \bibnamefont {Manalis}},\
  }\bibfield  {title} {\enquote {\bibinfo {title} {High-speed multiple-mode
  mass-sensing resolves dynamic nanoscale mass distributions},}\ }\href
  {https://www.nature.com/articles/ncomms8070} {\bibfield  {journal} {\bibinfo
  {journal} {Nature Communications}\ }\textbf {\bibinfo {volume} {6}},\
  \bibinfo {pages} {7070} (\bibinfo {year} {2015})}\BibitemShut {NoStop}%
\end{thebibliography}%

\end{document}